\documentclass[a4paper,12pt]{article}

\usepackage{amsmath,amssymb}
\usepackage{graphics}
\usepackage {pstricks,pst-node,pst-coil}
\usepackage {latexsym,amssymb}
\usepackage{lmodern}
\usepackage{amsthm} %łatwiejszy skład matematyki
\usepackage{float}
\usepackage{setspace}    % zmiana interlinii (np. 1,5 wiersza odstępu)
\usepackage [T1]{fontenc}
\usepackage[utf8]{inputenc} %kodowanie UTF-8
\usepackage{pgfplots}
\usepackage{makecell}

\begin{document}
\title{Reducing of system of partial differential equations and generalized symmetry of ordinary differential equations}
\author{I. M. Tsyfra$^1$, P. Sitko$^1$}

\maketitle

%\tableofcontents

\textbf{Abstract.}

Symmetry reductions of systems of two nonlinear partial differential equations are studied.
We find ansatzes reducing system of partial differential equations to system of ordinary differential equations. 
The method is applied to system related to Korteweg -- de Vries (KdV) equation, and reaction--diffusion equations.
We have also shown the possibility of constructing  solution to  system of non-evolutionary equations (\ref{eq:PDE1_nieewolu1}), and  (\ref{eq8}), which contains one or two arbitrary functions.
\newline

A well-known method for constructing solutions to nonlinear differential equations, based on the generalized  conditional symmetry of differential equations, was proposed in the works~\cite{Fokas_Liu, Zhdanov1}. It allows reducing a partial differential equation to a system of ordinary differential equations and thus constructing its solutions. We use the approach of reduction proposed in~\cite{KamranOlverMilson, Svirshcevski} and their generalization~\cite{Tsyfra, TIRz}. In this context, it is necessary to mention the important works~\cite{samo02a, Sergey}. Nonlinear evolutionary equation systems related to the KdV equation, non-evolutionary equation systems, and reaction-diffusion equation systems are studied  in the article.

Consider the system of ordinary differential equations.
\begin{equation}
\label{eq:uklad_rownan_ODE_dotwierdzenia1}
\begin{cases}
u_{xx}-a(t,x)u=0,\\
v_{xx}-a(t,x)v=0,
\end{cases}
\end{equation}
where $a(t,x)$ is a function on $t,x$.  Following reference~\cite{TIRz}, we study the generalized symmetry of system (\ref{eq:uklad_rownan_ODE_dotwierdzenia1}).
We prove that system (\ref{eq:uklad_rownan_ODE_dotwierdzenia1}) admits the following generalized vector field
\begin{equation}
\label{eq:uklad_rownan_polewektorowe_dotwierdzenia1}
\begin{split}
Q&=\Bigl(u_t-\frac{1}{2}(3\frac{u_xu_{xx}}{u}-u_{xxx}-\frac{uv_{xxx}}{v}+2\frac{u_xv_{xx}}{v}+\frac{uv_xv_{xx}}{v^2})\Bigl)\partial_u+\\
&+\Bigl(v_t-\frac{1}{2}(3\frac{v_xv_{xx}}{v}-v_{xxx}-\frac{vu_{xxx}}{u}+2\frac{v_xu_{xx}}{u}+\frac{vu_xu_{xx}}{u^2})\Bigl)\partial_v,
\end{split}
\end{equation}
if and only if $a(t,x)$ satisfies the KdV equation
\begin{equation}
\label{eq:uklad_rownan_rownaniekdva_dotwierdzenia1}
a_t=6aa_x-a_{xxx}.
\end{equation}
It means that the infinitesimal criterion of invariance~\cite{Bluman_Kumei}
 \begin{equation*}
Q^{(2)}\Bigl(u_{xx}-a(t,x)u\Bigl)\Big|_{[K]=0}\equiv 0,
\end{equation*}
 \begin{equation*}
Q^{(2)}\Bigl(v_{xx}-a(t,x)v\Bigl)\Big|_{[K]=0}\equiv 0,
\end{equation*}
where $Q^{(2)}$ is the second prolongation of the generalized vector field $Q$, and $K$  denotes the differential consequences of the system of equations (\ref{eq:uklad_rownan_ODE_dotwierdzenia1}) with respect to the variables  $t, x$, is satisfied in this case.
It should be noted that the function included in the system (\ref{eq:uklad_rownan_ODE_dotwierdzenia1}) is an arbitrary particular solution of the KdV equation (\ref{eq:uklad_rownan_rownaniekdva_dotwierdzenia1}). If we succeed in constructing the general solution of system (\ref{eq:uklad_rownan_ODE_dotwierdzenia1}), we will derive an ansatz that reduces the system of evolutionary equations
%\label{eq:uklad_rownan_dotwierdzenia1}
\begin{equation}\label{sys08}
\begin{cases}
u_t=\frac{1}{2}\left (3\frac{u_xu_{xx}}{u}-u_{xxx}-\frac{uv_{xxx}}{v}+2\frac{u_xv_{xx}}{v}+\frac{uv_xv_{xx}}{v^2}\right ),\\
v_t=\frac{1}{2}\left (3\frac{v_xv_{xx}}{v}-v_{xxx}-\frac{vu_{xxx}}{u}+2\frac{v_xu_{xx}}{u}+\frac{vu_xu_{xx}}{u^2}\right ),
\end{cases}
\end{equation}
to a system of four ordinary differential equations. Note that system (\ref{sys08}) is invariant with respect to the Lie transformation groups with infinitesimal generators
 $Q_1=u\partial_u+u\partial_v$, and  $Q_2=v\partial_u+v\partial_v$. Then the system of equations (\ref{sys08}) is modified and takes the form
 \begin{equation}\label{sys98}
\begin{cases}
u_t=\frac{1}{2}\left (3\frac{u_xu_{xx}}{u}-u_{xxx}-\frac{uv_{xxx}}{v}+2\frac{u_xv_{xx}}{v}+\frac{uv_xv_{xx}}{v^2}\right )+\lambda_1 u+\lambda_2 v,\\
v_t=\frac{1}{2}\left (3\frac{v_xv_{xx}}{v}-v_{xxx}-\frac{vu_{xxx}}{u}+2\frac{v_xu_{xx}}{u}+\frac{vu_xu_{xx}}{u^2}\right )+\lambda_1 u+\lambda_2 v.
\end{cases}
\end{equation}

Given the specific form of the function $a(t,x)$, one
 can find an ansatz that is a solution to the system (\ref{eq:uklad_rownan_ODE_dotwierdzenia1}). The variable $t$  in (\ref{eq:uklad_rownan_ODE_dotwierdzenia1}), is then treated as a parameter. Let us take the stationary solution  $a(t,x)=\frac{2}{x^2}$ of equation (\ref{eq:uklad_rownan_rownaniekdva_dotwierdzenia1}). Then system (\ref{eq:uklad_rownan_ODE_dotwierdzenia1}) is integrable by quadratures and the general solution has the form
 \begin{equation}
\label{eq:uklad_rownan_ansatze_dotwierdzenia1}
\begin{cases}
u(t,x)=c_1(t)x^2+\frac{c_2(t)}{x},\\
v(t,x)=c_3(t)x^2+\frac{c_4(t)}{x}.
\end{cases}
\end{equation}

  Using the obtained ansatz, one can then find the system of reduced equations of system (\ref{sys98}). This is a system of ordinary differential equations
 \begin{equation}\label{redsyskdf}
\begin{cases}
c_1'(t)=\frac{1}{2}(\lambda_1c_1(t)+\lambda_2c_3(t)),\\
c_2'(t)=\frac{1}{2}(24c_1(t)+\lambda_1 c_2(t)+\lambda_2 c_4(t)),\\
c_3'(t)=\frac{1}{2}(\lambda_1c_1(t)+\lambda_2c_3(t)),\\
c_4'(t)=\frac{1}{2}(24c_1(t)+\lambda_1 c_2(t)+\lambda_2 c_4(t)).
\end{cases}
\end{equation}
By integrating the reduced system (\ref{redsyskdf})  and using (\ref{eq:uklad_rownan_ansatze_dotwierdzenia1}), we obtain the solution of system of equations (\ref{sys98})
\begin{equation*}
\begin{split}
u(t,x)=&(s_1+s_2e^{\frac{1}{2}(\lambda_1+\lambda_2)t})x^2+\\
&+\frac{1}{x(\lambda_1+\lambda_2)}(12s_2\lambda_1te^{\frac{1}{2}(\lambda_1+\lambda_2)t}+12s_2\lambda_2te^{\frac{1}{2}(\lambda_1+\lambda_2)t}+12s_1\lambda_1t+12s_1\lambda_2t+\\
&+2s_3e^{\frac{1}{2}(\lambda_1+\lambda_2)t}-24s_2e^{\frac{1}{2}(\lambda_1+\lambda_2)t}+s_4\lambda_1+s_4\lambda_4),\\
v(t,x)=&\frac{1}{\lambda_2}(s_2\lambda_2e^{\frac{1}{2}(\lambda_1+\lambda_2)t}-s_1\lambda_1)x^2+\\
&+\frac{1}{\lambda_2x(\lambda_1+\lambda_2)}(12s_2\lambda_1\lambda_2te^{\frac{1}{2}(\lambda_1+\lambda_2)t}+12s_2\lambda_2^2te^{\frac{1}{2}(\lambda_1+\lambda_2)t}-12s_1\lambda_1^2t\\
&-12s_1\lambda_1\lambda_2t++2s_3\lambda_2e^{\frac{1}{2}(\lambda_1+\lambda_2)t}-24s_2\lambda_2e^{\frac{1}{2}(\lambda_1+\lambda_2)t}-s_4\lambda_1^2-s_4\lambda_1\lambda_2),
\end{split}
\end{equation*}
where  $s_1, s_2, s_3, s_4$ are real constants.

In what follows, we define   the system of ordinary differential equations~\cite{TIRz}
\begin{equation}
\begin{cases}
\label{eq:kontrast_ODE}
u_{xx}+u_x^2=0,\\
v_x=vu_x.
\end{cases}
\end{equation}
The second equation in (\ref{eq:kontrast_ODE}) was selected to demonstrate the method's effectiveness.
We show that system (\ref{eq:kontrast_ODE}) possesses generalized symmetry given by vector field
\begin{equation*}
Q_{ne1}=\Bigl(v^2u_{xt}-F(u-\ln{v})\Bigl)\partial_u+\Bigl(vv_{xt}-e^uH(u-\ln{v})\Bigl)\partial_v,
\end{equation*}
so the conditions hold
\begin{equation*}
Q_{ne1}^{(2)}(u_{xx}+u_x^2)\Big|_{[K]=0}\equiv 0
\end{equation*}
and
 \begin{equation*}
Q_{ne1}^{(2)}(v_x-vu_x)\Big|_{[K]=0}\equiv 0,
\end{equation*}
where  $K$  denotes the differential consequences of the system of equations (\ref{eq:kontrast_ODE}) with respect to the variables  $t, x$.
Then,  the  example of a non--evolutionary system to which the reduction method can be applied is the system
\begin{equation}
\begin{cases}
\label{eq:PDE1_nieewolu1}
u_{xt}=\frac{1}{v^2}F(u-\ln{v}),\\
v_{xt}=\frac{e^u}{v}H(u-\ln{v}).
\end{cases}
\end{equation}
The ansatz reducing system (\ref{eq:PDE1_nieewolu1}) is obtained by solving the system of ordinary equations (\ref{eq:kontrast_ODE}),
and it has the form
\begin{equation}
\begin{cases}
\label{eq:kontrast_ansatze}
u(t,x)=\ln{(x+c_1(t))}+c_2(t),\\
v(t,x)=c_3(t)(x+c_1(t)).
\end{cases}
\end{equation}
System  (\ref{eq:PDE1_nieewolu1}) is reduced to a system of ordinary differential equations
\begin{equation*}
\begin{cases}
-c_3^2(t)c_1'(t)=F(c_2(t)-\ln{c_3(t)}),\\
c_3(t)c'_3(t)=e^{c_2(t)}H(c_2(t)-\ln{c_3(t)})
\end{cases}
\end{equation*}
 by virtue of the ansatz (\ref{eq:kontrast_ansatze}). Thus, we obtained a system of two equations with three unknown functions $c_1, c_2, c_3$. Obviously, the solution will then depend on one arbitrary  function on variable $t$.

We show also that the generelized vector field is admmited by system of ordinary differential equations (\ref{eq:kontrast_ODE})
\begin{equation*}
Q_{ne2}=\Bigl(vu_{xt}+v_xv_t-F(u-\ln{v})\Bigl)\partial_u+\Bigl(vv_{xt}-vH(u-\ln{v})\Bigl)\partial_v.
\end{equation*}
From this it follows that ansatz (\ref{eq:kontrast_ansatze}) reduces non--evolutionary system
\begin{equation}
\begin{cases}
\label{sysweq}
u_{xt}=\frac{1}{v}F(u-\ln{v})- \frac{u_xu_t}{v},\\
v_{xt}=H(u-\ln{v}),
\end{cases}
\end{equation}
to the system of  ordinary differential equations. Indeed, substituting (\ref{eq:kontrast_ansatze}) into (\ref{sysweq}) gives
\begin{equation}
\begin{cases}
\label{redeqs}
c'_3(t)=F(c_2(t)-\ln{c_3(t)}),\\
c'_3(t)=H(c_2(t)-\ln{c_3(t)}).
\end{cases}
\end{equation}
Thus, we obtained an inconsistent system of two ordinary differential equations for three unknown functions $c_1, c_2, c_3$ in the case when $F\neq H$.
We see that the generalized symmetry ensures a reduction to a system of ordinary differential equations, but does not guarantee the consistency of the reduced system.
System (\ref{eq:kontrast_ansatze}) is consistent if and only if $F = H$. In this case, the system (\ref{sysweq}) takes the form.
\begin{equation}
\begin{cases}
\label{eq8}
u_{xt}=\frac{1}{v}F(u-\ln{v})- \frac{u_xv_t}{v},\\
v_{xt}=F(u-\ln{v}).
\end{cases}
\end{equation}
Ansatz (\ref{eq:kontrast_ansatze}) reduces this system  to the  single ordinary differential equation
\begin{equation}\label{singeq}
c'_3(t)=F(c_2(t)-\ln{c_3(t)}).
\end{equation}
Then we conclude that the solution of (\ref{eq8}) will depend on two arbitrary functions on variable $t$.

Next, the system of partial differential equations is considered~\cite{BarJuryk1}
\begin{equation}
\label{eq:PDE_2_zmienne_zalezne}
\begin{cases}
u_t=(F_1(u,v)u_{x})_x+G_1(u,v)u_x+H_1(u,v),\\
v_t=(F_2(u,v)v_{x})_x+G_2(u,v)v_x+H_2(u,v).
\end{cases}
\end{equation}
The desired vector field is defined in the form of
\begin{equation}
\label{eq:operator_2_zmienne_zal}
\begin{split}
Q&=[(F_1(u,v)u_{x})_x+G_1(u,v)u_x+H_1(u,v)]\partial_u+\\
&+[(F_2(u,v)v_{x})_x+G_2(u,v)v_x+H_2(u,v)]\partial_v.
\end{split}
\end{equation}
Let us also consider a system of ordinary differential equations
\begin{equation}\label{syseq18}
\begin{cases}
u_{xx}=\frac{1}{u}u_x^2,\\
v_{xx}=\frac{1}{v}v_x^2.
\end{cases}
\end{equation}

We prove that system (\ref{syseq18}) admits the generalized vector field  (\ref{eq:operator_2_zmienne_zal}) if and only if
\begin{equation}
\begin{cases}
F_1(u,v)=(\lambda_{11}\ln{u}+\delta_{11}\ln{v}+\mu_{11}+\frac{s_{11}}{u})u_x,\\
G_1(u,v)=(\lambda_{21}\ln{u}+\delta_{21}\ln{v}+\mu_{21}),\\
H_1(u,v)=u(\delta_{31}\ln{v}+\lambda_{31}\ln{u}+\mu_{31}),\\
F_2(u,v)=(\lambda_{12}\ln{u}+\delta_{12}\ln{v}+\mu_{12}+\frac{s_{12}}{v})v_x,\\
G_2(u,v)=(\lambda_{22}\ln{u}+\delta_{22}\ln{v}+\mu_{22}),\\
H_2(u,v)=v(\delta_{32}\ln{v}+\lambda_{32}\ln{u}+\mu_{32}).\\
\end{cases}
\end{equation}
Then we obtain the following system of evolution equations
\begin{equation}\label{sysred}
\begin{cases}
u_t=[(\lambda_{11}\ln{u}+\delta_{11}\ln{v}+\mu_{11}+\frac{s_{11}}{u})u_x]_x+(\lambda_{21}\ln{u}+\delta_{21}\ln{v}+\mu_{21})u_x+\\
+u(\delta_{31}\ln{v}+\lambda_{31}\ln{u}+\mu_{31}),\\
v_t=[(\lambda_{12}\ln{u}+\delta_{12}\ln{v}+\mu_{12}+\frac{s_{12}}{v})v_x]_x+(\lambda_{22}\ln{u}+\delta_{22}\ln{v}+\mu_{22})v_x+\\
+v(\delta_{32}\ln{v}+\lambda_{32}\ln{u}+\mu_{32}).
\end{cases}
\end{equation}
Solving system of ordinary differential equations (\ref{syseq18}) we obtain the corresponding ansatz
\begin{equation}\label{anz08}
u(t,x)=c_2(t)e^{c_1(t)x},\quad v(t,x)=c_4(t)e^{c_3(t)x}.
\end{equation}
Substituting (\ref{anz08}) into (\ref{sysred}) we obtain reduced system of ordinary differential equations
\begin{equation*}
\begin{cases}
c_1'=\lambda_{11}c_1^3+\delta_{11}c_1^2c_3+\lambda_{21}c_1^2+\delta_{21}c_1c_3+\lambda_{31}c_1+\delta_{32}c_3,\\
c_2'=c_2(\lambda_{11}c_1^2\ln{c_2}+\delta_{11}c_1^2\ln{c_4}+\lambda_{21}c_1\ln{c_2}+\delta_{21}c_1\ln{c_4}+\delta_{11}c_1c_3+\lambda_{11}c_1^2+\\+\mu_{11}c_1^2+\lambda_{31}\ln{c_2}+\delta_{32}\ln{c_4}+\mu_{21}c_1+\mu_{31}),\\
c_3'=\lambda_{12}c_1c_3^2+\delta_{12}c_3^3+\lambda_{22}c_1c_3+\delta_{22}c_3^2+\lambda_{32}c_1+\delta_{32}c_3,\\
c_4'=c_4(\lambda_{12}c_3^2\ln{c_2}+\delta_{12}c_3^2\ln{c_4}+\lambda_{22}c_3\ln{c_2}+\delta_{22}c_3\ln{c_4}+\lambda_{12}c_1c_3+\delta_{12}c_3^2+\\+\mu_{12}c_3^2+\lambda_{32}\ln{c_2}+\delta_{32}\ln{c_4}+\mu_{22}c_3+\mu_{32}).
\end{cases}
\end{equation*}

We show the application of the inverse symmetry reuction method for reducting systems of two partial  differential equations to the systems of ordinary differential equations. We show that the system of evolutionary equations (\ref{sys98}) can be reduced to a system of ordinary equations, parameterized by an arbitrary solution of the KdV equation.
We have also demonstrate the possibility of constructing the solution to the system of non--evolutionary equations (\ref{eq:PDE1_nieewolu1}), and  (\ref{eq8}) which contains one  or two arbitrary functions.

$^1$\noindent Ivan Tsyfra\\  %Name Surname
tsyfra@agh.edu.pl\\
ORCID: https://orcid.org/0000-0001-6665-3934 \bigskip % ORCID identifier (optional)

\noindent {\small
\noindent AGH University of Krakow \\
 Faculty of Applied Mathematics\\
30 Mickiewicza Avenue, 30-059, Krakow, Poland.
}\bigskip

$^1$\noindent Patryk Sitko\\  %Name Surname
psitko@agh.edu.pl\\
ORCID: https://orcid.org/0000-0002-2510-6528 \bigskip % ORCID identifier (optional)

\noindent {\small
\noindent AGH University of Krakow\\
Faculty of Applied Mathematics\\
30 Mickiewicza Avenue, 30-059, Krakow, Poland.
}\bigskip


\begin{thebibliography}{88}

\bibitem{BarJuryk1} A. F. Barannyk, T. A. Barannyk, and I. I. Yuryk, \textit{A method for the construction of exact solutions to the nonlinear heat equation $U_T=(F(U)U_X)_X+G(U)U_X+H(U)$}, Ukrainian Mathematical Journal, Vol. 71, No. 11, April, 2020 (Ukrainian Original Vol. 71, No. 11, November, 2019).

\bibitem{Bluman_Kumei} G. W. Bluman, S. Kumei, \textit{Symmetries and differential equations}, Springer-Verlag,New York, 1981.

\bibitem{Fokas_Liu} A. S. Fokas, Q. M. Liu, \textit{Nonlinear interaction of traveling waves of non-integrable equations}, Phys.
Rev. Lett., \textbf{72} (1994), 3293–3296. https://doi.org/10.1103/PhysRevLett.72.3293

\bibitem{KamranOlverMilson} N. Kamran, R. Milson, P. J. Olver, \textit{Invariant Moduler and the Reduction of Nonlinear Partial Differential Equations to Dynamical Systems}, Advances in Mathematics, \textbf{156}, 286-319 (2000).

\bibitem{samo02a} A. Samokhin, \textit{Full Symmetry Algebra for ODEs and Control Systems}, Acta Applicandae Mathematicae 72: 87-99, 2002.

\bibitem{Sergey} A. Sergyeyev, \textit{Constructing conditionally integrable evolution systems in (1+1) dimensions: a generalization of invariant modules approach}, J. Phys. A: Math. Gen. \textbf{35} (2002) 7653-7660.

\bibitem{Svirshcevski} S. R. Svirshchevskii, \textit{Lie-B\"{a}cklund symmetries of linear ODEs and generalized separation of variables in nonlinear equations}, Physics Letters A 199 (1995) 344-348.

\bibitem{Tsyfra} I. Tsyfra, \textit{On the Symmetry Approach to Reduction of Partial Differential Equations}, Proceeding of Institute of Mathematics of NAS of Ukraine, 2004.

\bibitem{TIRz}
Ivan Tsyfra, Wojciech Rzeszut
\textit{On reducing and finding solutions of nonlinear evolutionary equations via generalized symmetry of ordinary differential equation},
MBE, 19(7): 6962--6984, DOI: 10.3934/mbe.2022328

\bibitem{Zhdanov1}
R. Z. Zhdanov, \textit{Conditional Lie–Bäcklund symmetry and reduction of evolution equations}, J. Phys.
 A: Math. Gen., \textbf{28} (1995), 3841–3850. https://doi.org/10.1088/0305-4470/28/13/027.

\end{thebibliography}
\end{document}